\documentclass{iaus}
\usepackage{graphicx}

\title[SED for H\,{\sc ii} region in M33] 
{Spectral Energy Distributions of a set of H\,{\sc ii} regions in M33 (HerM33es)}

\author[M. Rela\~no \& the HerM33es Team]   
{M. Rela\~no$^1$,
 S. Verley$^1$,
 I. P\'erez$^1$,
 C. Kramer$^2$,
 E. M. Xilouris$^3$,
 M. Boquien$^4$, 
 J. Braine$^5$,
 D. Calzetti$^6$,
 C. Henkel$^7$,  
 \and HerM33es Team}
\affiliation{$^1$Dept. de F\'\i sica Te\'orica y del Cosmos, Universidad de Granada, Spain,
$^2$Instituto de Radioastronom\'\i a Milim\'etrica, Granada, Spain,
$^3$Institute of Astronomy and Astrophysics, NOA, Athens, Greece,
$^4$Laboratoire d'Astrophysique de Marseille, Marseille, France,
$^5$Laboratoire d'Astrophysique de Bordeaux, France,
$^6$Deparment of Astronomy, University of Massachusetts, USA,
$^7$MPI f\"ur Radioastronomie, Bonn, Germany
}

\pubyear{2011} 
\volume{284}  
\pagerange{1--12}
\jname{The Spectral Energy Distribution of Galaxies}
\editors{R.J. Tuffs \&  C.C.Popescu, eds.}

\newcommand{\mi}{\relax \ifmmode {\mu{\mbox m}}\else $\mu$m\fi}
\newcommand{\hii}{\relax \ifmmode {\mbox H\,{\sc ii}}\else H\,{\sc ii}\fi}
\newcommand{\sii}{\relax \ifmmode {\mbox S\,{\scshape ii}}\else S\,{\scshape ii}\fi}
\newcommand{\nii}{\relax \ifmmode {\mbox N\,{\scshape ii}}\else N\,{\scshape ii}\fi}
\newcommand{\neii}{\relax \ifmmode {\mbox Ne\,{\scshape ii}}\else Ne\,{\scshape ii}\fi}
\newcommand{\neiii}{\relax \ifmmode {\mbox Ne\,{\scshape iii}}\else Ne\,{\scshape iii}\fi}
\newcommand{\oiii}{\relax \ifmmode {\mbox O\,{\scshape iii}}\else O\,{\scshape iii}\fi}
\newcommand{\oii}{\relax \ifmmode {\mbox O\,{\scshape ii}}\else O\,{\scshape ii}\fi}
\newcommand{\oi}{\relax \ifmmode {\mbox O\,{\scshape i}}\else O\,{\scshape i}\fi}
\newcommand{\ha}{\relax \ifmmode {\mbox H}\alpha\else H$\alpha$\fi}

\newcommand{\hep}{\relax \ifmmode {\mbox H}\epsilon\else H$\epsilon$\fi}
\newcommand{\hdel}{\relax \ifmmode {\mbox H}\delta\else H$\delta$\fi}
\newcommand{\hgam}{\relax \ifmmode {\mbox H}\gamma\else H$\gamma$\fi}

\newcommand{\pa}{\relax \ifmmode {\mbox Pa}\alpha\else Pa$\alpha$\fi}
\newcommand{\hb}{\relax \ifmmode {\mbox H}\beta\else H$\beta$\fi}
\newcommand{\rdostres}{\relax \ifmmode {\,\mbox{R}}_{\rm 23}\else \,\mbox{R}$_{\rm 23}$\fi}
\newcommand{\ergs}{\relax \ifmmode {\,\mbox{erg\,s}}^{-1}\else \,\mbox{erg\,s}$^{-1}$\fi}
\newcommand{\me}{\relax \ifmmode {\,}^{-1}\else \,$^{-1}$\fi}

\newcommand{\msun}{\relax \ifmmode {\,\mbox{M}}_{\odot}\else \,\mbox{M}$_{\odot}$\fi}

\newcommand{\cmtres}{\relax \ifmmode {\,\mbox{cm}}^{-3}\else \,\mbox{cm}$^{-3}$\fi}
\newcommand{\cmdos}{\relax \ifmmode {\,\mbox{cm}}^{-2}\else \,\mbox{cm}$^{-2}$\fi}
\newcommand{\cmseis}{\relax \ifmmode {\,\mbox{cm}}^{-6}\else \,\mbox{cm}$^{-6}$\fi}
\newcommand{\hi}{\relax \ifmmode {\mbox H\,{\scshape i}}\else H\,{\scshape i}\fi}

\begin{document}

\maketitle

\begin{abstract}
Within the framework of the HerM33es Key Project for Herschel and in combination with multi-wavelength data, we study the Spectral Energy Distribution (SED) of a set of \hii\ regions in the Local Group Galaxy M33. Using the \ha\ emission, we perform a classification of a selected \hii\ region sample in terms of morphology, separating the objects in {\it filled}, {\it mixed}, {\it shell} and {\it clear shell} objects. We obtain the SED for each \hii\ region as well as a representative SED for each class of objects. We also study the emission distribution of each band within the regions. We find different trends in the SEDs for each morphological type that are related to properties of the dust and their associated stellar cluster. The emission distribution of each band within the region is different for each morphological type of object. 
\end{abstract}

\firstsection 
\section{Introduction}
The study of the star formation rate (SFR) in galaxies of different types has been lately improved due to the new available data. Remarkably, the focus has been turned to the closest galaxies 
as well as to star-forming regions within our Galaxy. For these nearby objects, the resolution of the data offers us an opportunity to test 
whether the proposed SFR calibrators trace indeed the location of the stellar births (Churchwell et al. 2006; Rela\~no \& Kennicutt 2009, among others). 
Recently, a new study on the star-forming regions in the Magellanic Clouds has analysed the relation of the amount of flux at the different wavelengths via the SEDs of these objects (Lawton et al. 2010). 

Within the HerM33es Key Project (Kramer et al. 2010) we are obtaining maps of the entire galaxy M33 at 
wavelengths between 100\,\mi\ and 500\,\mi\ using PACS and SPIRE instruments on Herschel. Verley et al (2010) identified a set of \hii\ regions in the north of M33 showing a shell-like morphology in these infrared bands, and also traced by the \ha\ emission. In order to further study this phenomenon we have analysed the SEDs of a set of \hii\ regions in M33 covering different morphologies. The resolution of our data (from $\sim2''$ to $\sim20^{\prime\prime}$) is good enough to perform such as study. 

\section{SED of H\,{\sc ii} regions}
We use an \ha\ image of M33 (Hoopes \& Walterbos 2000) to identify a set of \hii\ regions for which 
clear morphology can be recognised. A morphological classification was obtained with the following criteria: 
{\it filled} regions are objects showing a compact knot, {\it mixed} regions 
are those presenting several compact knots and filamentary structures, 
and {\it shells} are regions showing arcs. We add another classification for 
the most spherical, closed shells called {\it clear shells}. Out of 
the 120 selected \hii\ regions, 9 are filled, 47 mixed, 37 shell and 27 clear shells. 
Our sample is distributed over the whole disk of M33 (see Fig.~1, left).

We use multi-wavelength data from FUV (GALEX) to 250\,\mi\ (Herschel) smoothed to a common 20$^{\prime\prime}$ resolution  and regridded to a 6$^{\prime\prime}$ pixel size (corresponding to those of the 250\,\mi\ Herschel image)  to obtain the SED for each region. Photometry was performed using individual apertures for each object and local background was subtracted to eliminate the contribution of the diffuse medium to the \hii\ region fluxes. In Fig.~1 (right) we show the SEDs for 
our objects together with a {\it characteristic} SED for each classification obtained with the mean values in each band for all the \hii\ regions in the corresponding classification. From Fig.~1 (right) we observe the following trends: (i) mixed regions are more luminous 
in all bands as they normally have several knots of star formation, (ii) the slope of the SED between the FUV-NUV wavelength range 
and \ha\ is steeper for shells and clear shells than for filled and mixed, (iii) filled and mixed objects have more 
24\,\mi\ relative to 8\,\mi\ than the shells and clear shells do.

   \begin{figure}
   \centering
  \includegraphics[width=0.45\textwidth]{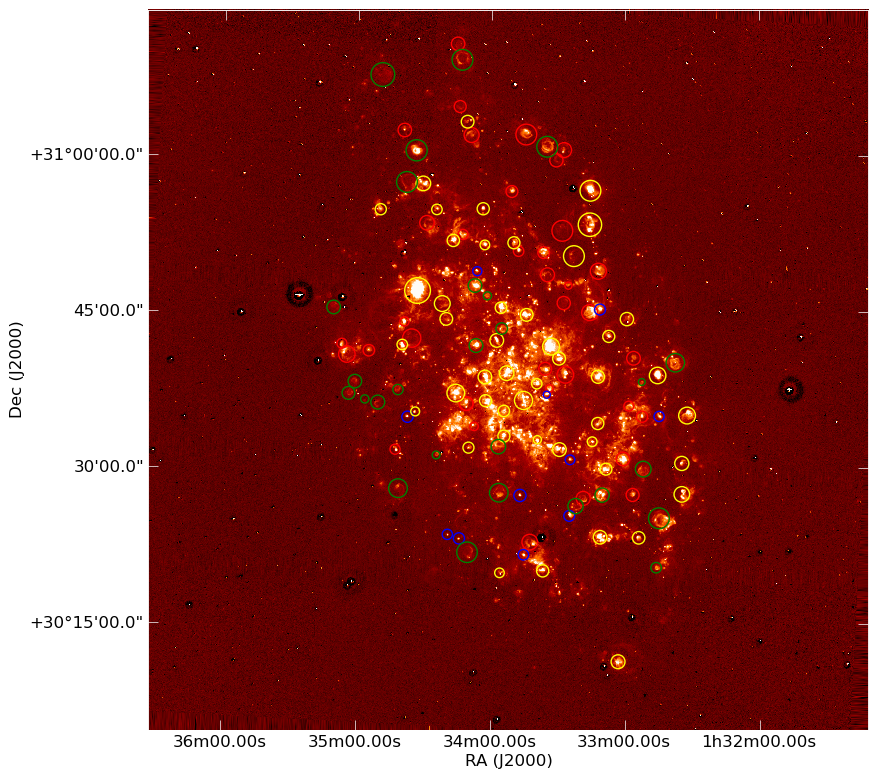}   
 \includegraphics[width=0.45\textwidth]{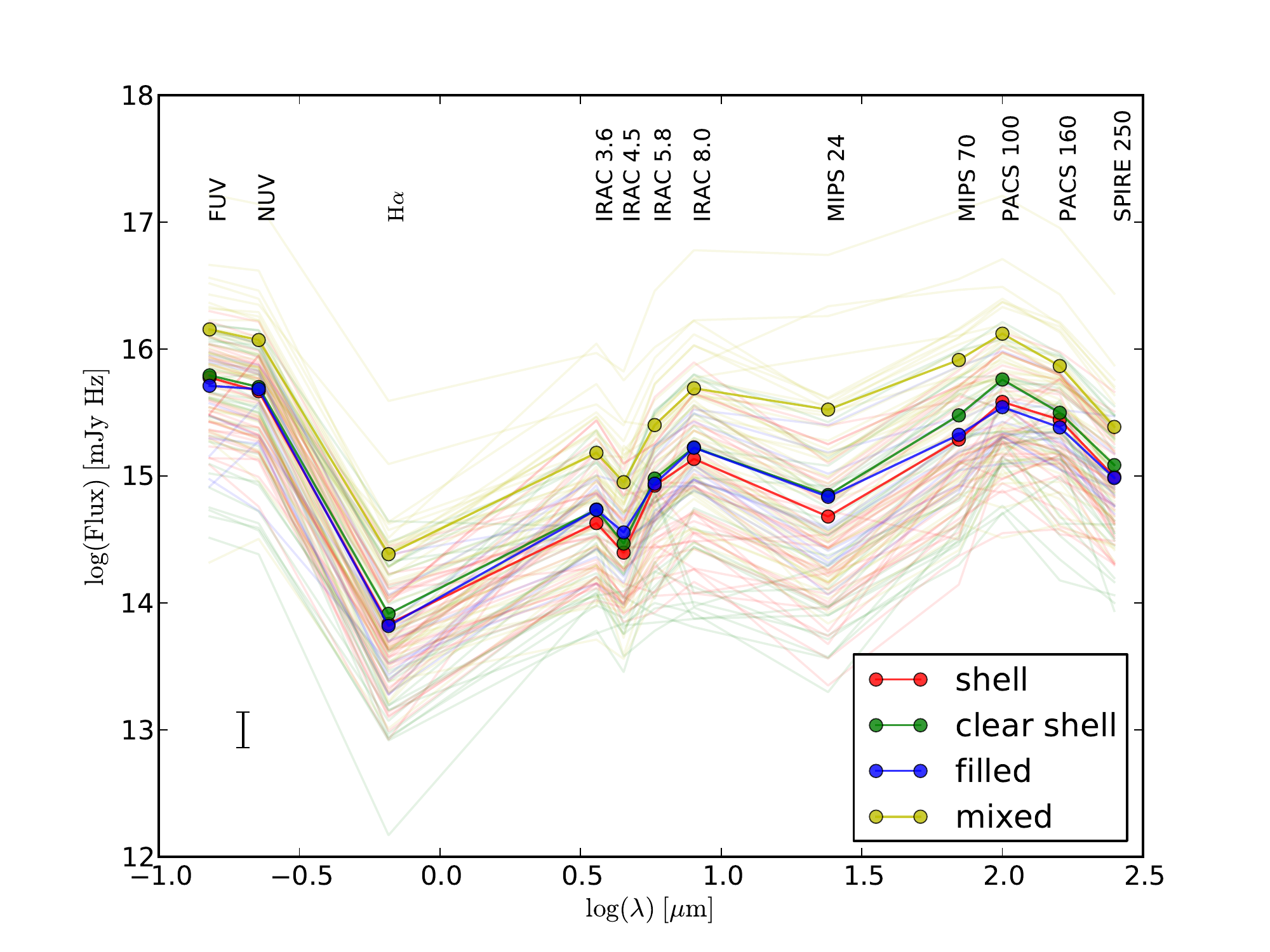}
   \caption{Left: Location of the \hii\ region sample on the continuum-subtracted \ha\ image of M33 (Hoopes \& Walterbos 2000). 
   Circles correspond to the apertures used to perform the photometry. Right: SED for our set of \hii\ regions. Typical errors for the fluxes are shown in the lower left corner of the figure. }
              \label{hiireg}%
    \end{figure}

\section{Dust Temperature}
The 100\,\mi/70\,\mi, 160\,\mi/70\,\mi\ or 160\,\mi/100\,\mi\ ratios normally trace the temperature of the warm dust emitting from 24\,\mi\ to 160\,\mi. In Fig.~2 (left) we plot the 100\,\mi/70\,\mi\ ratio versus the \ha\ surface brightness for our sample. At high \ha\ 
surface brightness the 100\,\mi/70\,\mi\ flux density ratio decreases showing that highly luminous \hii\ regions tend to have warmer dust. This agrees with the correlation observed by Boquien et al. (2010, 2011) in M33, and in a similar study by Bendo et al. (2011) who observed M81, M83 and NGC~2403 with Herschel. 

However, the filled regions seem to have a constant 100\,\mi/70\,\mi\ ratio, independent of the \ha\ surface brightness. For shells and clear shells there is a dispersion in the 100\,\mi/70\,\mi\ ratio showing that these regions present a range of dust temperatures. For the shells, the relative location between stars and dust can affect more the temperature of the dust than the intensity of the stellar radiation field and therefore they tend to present a wider dust temperature range.

   \begin{figure}
   \centering 
  \includegraphics[width=0.6\textwidth]{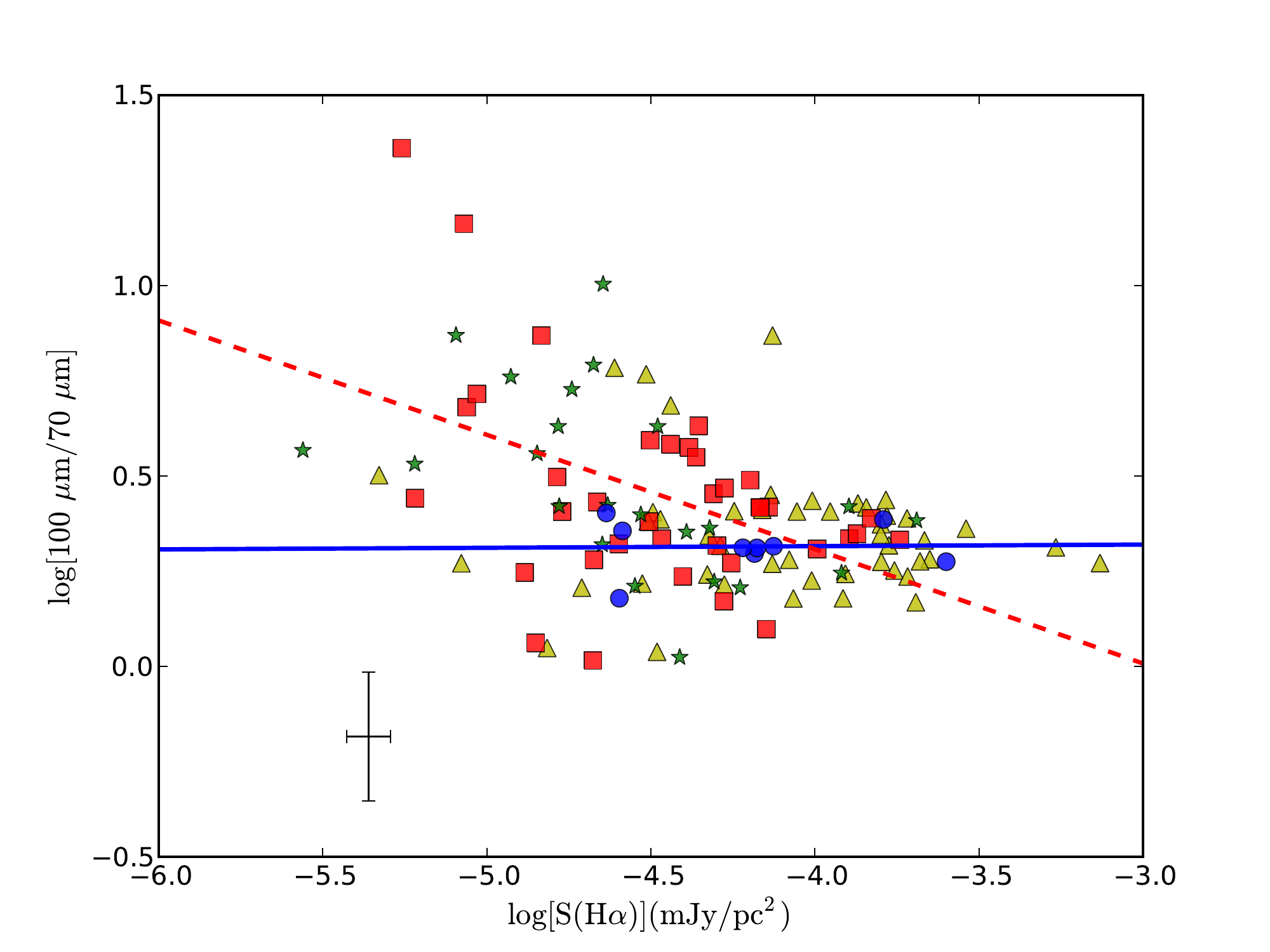}   
   \includegraphics[width=0.3\textwidth]{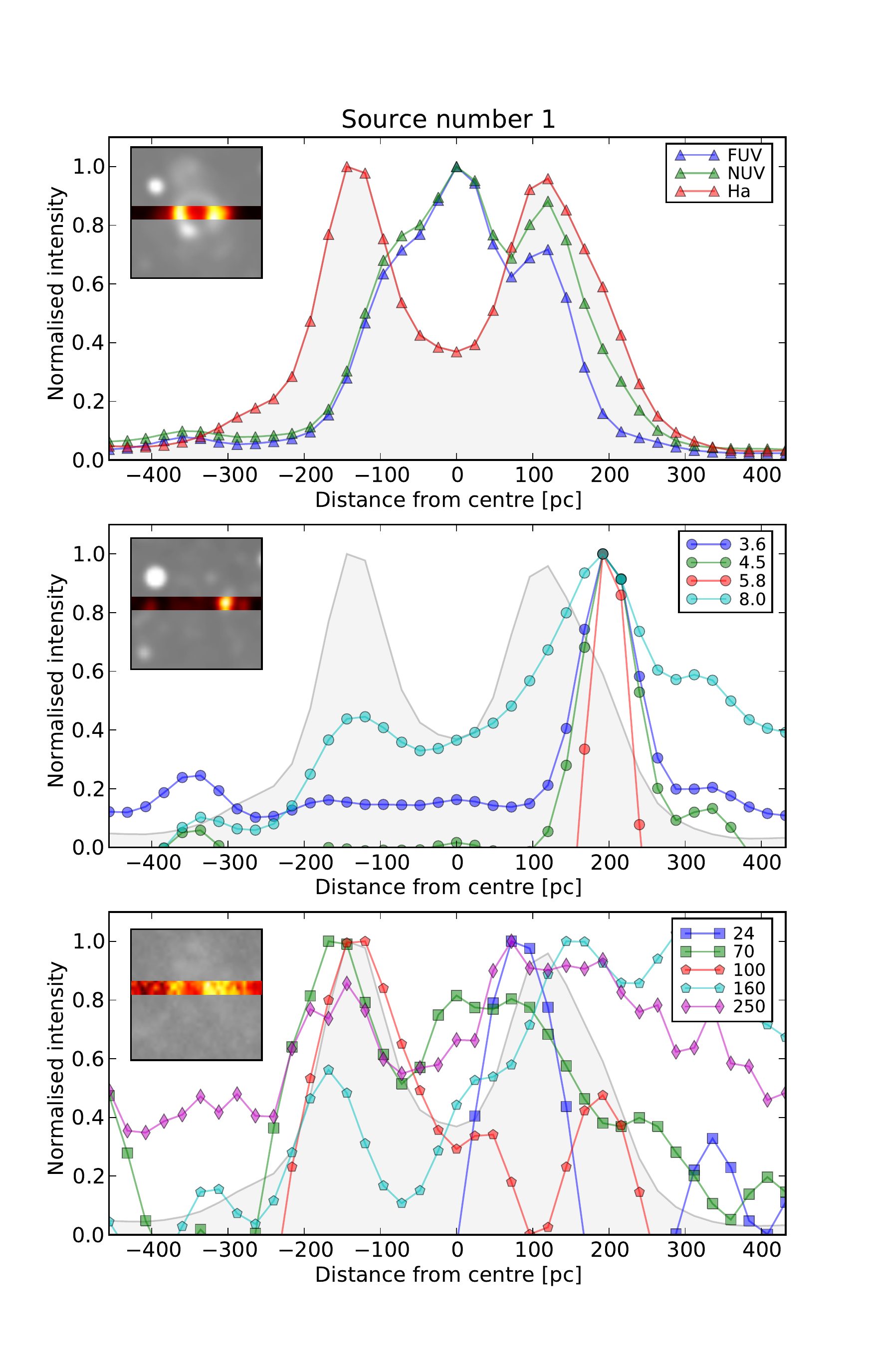} 
   \caption{Left: 100\,\mi/70\,\mi\ ratio versus the \ha\ surface brightness for our sample. Colour code is the same 
   as in Fig.1. Right: Normalised emission line profiles for a shell region in the horizontal direction. Profiles are separated in three panels (top: \ha, FUV, and NUV, middle: 3.6\,\mi, 4.5\,\mi, 5.8\,\mi, and 8.0\,\mi\ and bottom: 24\,\mi, 70\,\mi, 100\,\mi, 160\,\mi, and 250\,\mi). At top-left corner in each panel we show the location of the profile in the region in the \ha, 4.5\,\mi, and 250\,\mi\ images for the top, middle and bottom panels, respectively. The \ha\ profile is depicted in grey in all the panels for reference.}
              \label{Tdust}%
    \end{figure}
\section{Emission line profiles for clear shells}

We have performed a multi-wavelength study of the emission distribution in the interior of the clear shells. From FUV to 250\,\mi\ 
 we have obtained profiles in the horizontal (East-West) direction crossing the centre 
 of the \hii\ regions. Each profile corresponds to the integration of a line of 4 
 pixels ($\sim 24''$) width perpendicular to the direction of the profile. 
  
In Fig.~\ref{Tdust} (right) we show the emission line profile for one of the clear shells of our sample. 
The \ha\ profile shows the characteristic double peak of the shell emission, 
and there is a displacement between the \ha\ and FUV/NUV emission, with the FUV/NUV 
emission located in the inner part of the region. The ratio \ha/FUV is lower in the centre than in the boundaries of the shell, which could be due to: (i) the existence of a young stellar population within the shell, or (ii) the ionising photons from the central cluster reaching the shell and ionising the gas within the rim. The emission at all IR bands follows clearly the \ha\ shape of the shell: at 24\,\mi\ and 250\,\mi\ the emission decays 
in the centre of the shell and they are enhanced at the boundaries. The same trend, though not so clear, is seen at 70\,\mi,
100\,\mi, and 160\,\mi. The emission of the Polycyclic Aromatic Hydrocarbon molecules (PAH) at 8\,\mi\ is marked by the location of the shell boundaries.


\begin{thebibliography}{}

\bibitem[Bendo et al. (2011)]{}
{Bendo, G. J., Boselli, A., Dariush, A. et al.} 2011, 
\textit{astro-ph}, 1109.0237

\bibitem[Boquien et al. (2010)]{}
{Boquien, M., Calzetti, D., Kramer, C. et al. } 2010, 
\textit{A\&A}, 518, 70

\bibitem[Boquien et al. (2011)]{}
{Boquien, M., Calzetti, D., Combes, F. et al. } 2011, 
\textit{AJ}, 142, 111

\bibitem[Churchwell et al. (2006)]{}
{Churchwell, E., Povich, M. S., Allen, D. et al.} 2006, 
\textit{ApJ}, 649, 759

\bibitem[Hoopes \& Walterbos (2000)]{}
{Hoopes, C. G., Walterbos, R. A. M.} 2000, 
\textit{ApJ}, 541, 597


\bibitem[Kramer et al. (2010)]{}
{Kramer, C., Buchbender, C., Xilouris, E. M. et al. } 2010, 
\textit{A\&A}, 518, 67



\bibitem[Lawton et al. (2010)]{}
{Lawton, B., Gordon, K. D., Babler, B. et al. } 2009, 
\textit{ApJ}, 716, 453


\bibitem[Rela\~no, M. \& Kennicutt, R. C. Jr. (2009)]{}
{Rela\~no, M. \& Kennicutt, R. C. Jr.} 2009, 
\textit{ApJ}, 699, 1125

\bibitem[Verley et al. (2010)]{}
{Verley, S., Rela\~no, M., Kramer, C. et al. } 2010, 
\textit{A\&A}, 518, 68





\end{thebibliography}
\end{document}